\documentclass[conference]{IEEEtran}

\usepackage{amsmath,amssymb,amsthm}
\usepackage{mathtools}
\usepackage{bm}
\usepackage{graphicx}
\usepackage{xcolor}
\usepackage[normalem]{ulem}
\usepackage[colorlinks=true,allcolors=black]{hyperref}
\usepackage{cleveref}
\usepackage{orcidlink}
\usepackage{balance}
\usepackage{tikz}
\usetikzlibrary{
    arrows.meta,
    calc,
    patterns,
    positioning
}

\definecolor{tgtc}{HTML}{2A78D6}
\definecolor{cltc}{HTML}{EB6834}
\definecolor{sigc}{HTML}{6250D6}
\definecolor{inkc}{HTML}{52514E}

\crefname{figure}{Fig.}{Figs.}
\Crefname{figure}{Fig.}{Figs.}
\crefname{lemma}{Lemma}{Lemmas}
\Crefname{lemma}{Lemma}{Lemmas}
\crefname{corollary}{Corollary}{Corollaries}
\Crefname{corollary}{Corollary}{Corollaries}
\crefname{proposition}{Proposition}{Propositions}
\Crefname{proposition}{Proposition}{Propositions}
\crefname{theorem}{Theorem}{Theorems}
\Crefname{theorem}{Theorem}{Theorems}
\crefname{remark}{Remark}{Remarks}
\Crefname{remark}{Remark}{Remarks}
\crefname{section}{Section}{Sections}
\Crefname{section}{Section}{Sections}
\usepackage{pgfplots}
\pgfplotsset{compat=1.18}
\pgfplotstableset{col sep=comma}
\newtheorem{theorem}{Theorem}

\newcommand{\C}{\mathbb{C}}
\newcommand{\E}{\mathbb{E}}
\newcommand{\herm}{^{\mathsf{H}}}
\newcommand{\transp}{^{\mathsf{T}}}

\DeclareMathOperator{\tr}{tr}

\newcommand{\Rx}{\mathbf{R}_x}
\newcommand{\Cx}{\mathbf{C}_x}
\newcommand{\Rd}{\mathbf{R}_d}
\newcommand{\Rtd}{\widetilde{\mathbf{R}}_d}
\newcommand{\steer}{\mathbf{a}}
\newcommand{\CRB}{\mathrm{CRB}}

\begin{document}

\title{Improper Signaling for Clutter-Limited Angle Estimation in ISAC}

\author{
\IEEEauthorblockN{Victor~Augusto~Del~Monego\IEEEauthorrefmark{1}\textsuperscript{\orcidlink{0009-0001-6817-3093}}, Polianna~Basso\IEEEauthorrefmark{1}\textsuperscript{\orcidlink{0009-0000-2202-7464}}, Murilo~Capponi\IEEEauthorrefmark{1}\textsuperscript{\orcidlink{0009-0005-5740-9952}},
Bruno~S.~Chang\IEEEauthorrefmark{1}\textsuperscript{\orcidlink{0000-0003-0232-7640}},\\
Yahia Medjahdi\IEEEauthorrefmark{2}\textsuperscript{\orcidlink{0000-0001-7030-1130}},
Giuseppe~Thadeu~Freitas~de~Abreu\IEEEauthorrefmark{3}\textsuperscript{\orcidlink{0000-0002-5018-8174}},
and Didier~Le~Ruyet\IEEEauthorrefmark{4}\textsuperscript{\orcidlink{0000-0002-9673-2075}}}

\vspace{0.7em}

\IEEEauthorblockA{\IEEEauthorrefmark{1}CPGEI/Electronics Department, Federal University of Technology - Paran\'a, Curitiba, Brazil\\
\IEEEauthorrefmark{2} IMT Nord Europe, Institut Mines T\'el\'ecom, Center for Digital Systems, F-59653 Villeneuve d’Ascq, France\\
\IEEEauthorrefmark{3}School of Computer Science and Engineering, Constructor University, Bremen, Germany\\
\IEEEauthorrefmark{4}CEDRIC/Conservatoire National des Arts et M\'etiers - Paris, France\\[0.5ex]
Emails: \{poliannabasso, victormonego, murilocapponi\}@alunos.utfpr.edu.br, bschang@utfpr.edu.br,\\
yahia.medjahdi@imt-nord-europe.fr, gabreu@constructor.university, didier.le\_ruyet@cnam.fr}
}

\maketitle

\begin{abstract}

Integrated sensing and communication (ISAC) transmit designs typically optimize the transmit covariance under proper signaling, leaving the potential sensing benefits of pseudo-covariance unexplored. This paper investigates the conditions under which transmit pseudo-covariance can influence the angle Cramér–Rao bound (CRB) of a deterministic target in the presence of clutter. For a prescribed transmit covariance and a single clutter patch, we establish that CRB sensitivity occurs if and only if the clutter is illuminated, exhibits nonzero non-circularity, and overlaps the angle-informative component of the target response. This characterization demonstrates that angular separation alone does not determine whether transmit impropriety can improve sensing performance. We derive a closed-form optimal incident-field pseudo-variance that minimizes the angle CRB over the set of feasible transmit pseudo-covariances. Under fully non-circular clutter scattering, the optimal design eliminates the clutter-induced CRB penalty. Numerical results for a 32-antenna ISAC system show CRB reductions of up to 99\% relative to proper sensing under complete overlap and 30\% under partial overlap, as well as up to approximately 7~dB lower clutter-induced CRB penalty across target--clutter separations. These findings identify transmit pseudo-covariance as a conditional sensing design resource whose effectiveness depends on clutter statistics and projected target–clutter geometry.

\end{abstract}


\section{Introduction}
\label{sec:intro}

In integrated sensing and communication (ISAC), the transmitted waveform conveys information to the users while its echoes are used to estimate target parameters, such as the target angle, in the presence of disturbance~\cite{xiong2023tradeoff}. The transmit design is therefore determined not only by its average power distribution, but also by the second-order statistics of the waveform. The Hermitian covariance defines how the signal spatially allocates power but, by itself, does not completely describe these statistics: the complementary pseudo-covariance is also required, and it is zero for proper signaling \cite{schreier2010complex}.
This additional statistic is relevant for transmit design because it can be modified without changing the transmit beampattern, providing a degree of freedom beyond the covariance. Improper Gaussian signaling, which exploits it, has been shown to outperform proper signaling in interference-limited communication systems \cite{cadambe2010asymmetric,zeng2013igs,lameiro2017zifc,javed2020journey}. 
In ISAC, Zhu \emph{et al.}~\cite{zhu2026phased} use improper signaling in hybrid beamforming to balance multiuser rates against a transmit-beampattern sensing objective; at a prescribed covariance, that objective is unchanged by the pseudo-covariance.
For sensing, however, most Cram\'er--Rao bound (CRB)-based ISAC transmit designs restrict the signaling to the proper case by optimizing the covariance alone \cite{liu2022crb,hua2024mimoisac}. This raises the question of whether the transmit pseudo-covariance can affect the sensing bound and, if so, under which conditions and how it should be chosen.

This question is related to recent ISAC results on CRB-rate tradeoffs, although the mechanism considered here is different. In \cite{song2025crbrate}, the sensing-communication tradeoff is studied using superimposed Gaussian information and deterministic sensing signals. Under a different system model, \cite{guo2026crbrate} establishes that circularly symmetric input distributions achieve the CRB-rate Pareto boundary. Here, the clutter is signal-dependent and possibly non-circular, which allows the transmit pseudo-covariance to affect the disturbance at the sensing receiver without a separate sensing signal. Non-circularity has also been exploited in direction finding, mainly as a property of the received signal or through receiver-side widely linear processing \cite{delmas2004nccrb,abeida2017rectilinear,steinwandt2016nccrb,picinbono1995wl,chevalier2006saic}, and recent ISAC works have considered more general second-order input statistics as sensing variables \cite{lu2024randomisac,du2024reshaping}. The focus here is specifically on the transmit pseudo-covariance and the deterministic-target angle CRB.

The paper investigates improper Gaussian signaling as a transmit-side resource for ISAC sensing. The transmit pseudo-covariance does not change the target echo; instead, it acts indirectly, by shaping the pseudo-covariance of the signal-dependent clutter observed by a widely linear receiver. For a single clutter patch, it is shown that the pseudo-covariance affects the angle CRB if and only if the clutter is illuminated, non-circular, and overlaps the angle-informative component of the target response; angular separation alone is not sufficient. Under these conditions, the optimal pseudo-covariance and resulting CRB reduction are obtained in closed form, without numerical optimization: the optimal design is maximally improper along the clutter direction, with a phase set by the clutter non-circularity and the target--clutter geometry.  Numerical results for a 32-element array show reductions of up to approximately 7~dB in the clutter-induced CRB penalty relative to proper signaling.

The remainder of this paper is organized as follows. Section~\ref{sec:model} presents the signal model and the augmented angle CRB. Section~\ref{sec:conditions} establishes the conditions for pseudo-covariance sensitivity and derives the closed-form optimal design. Section~\ref{sec:validations} presents the numerical results, and Section~\ref{sec:conc} concludes the paper.

\emph{Notation.} Boldface lowercase and uppercase letters denote vectors and matrices,
respectively. The operators $(\cdot)^\ast$, $(\cdot)\transp$, and $(\cdot)\herm$ denote complex
conjugation, transpose, and Hermitian transpose, respectively. The expectation operator is denoted by $\E[\cdot]$, and $\Re(\cdot)$ and $\Im(\cdot)$ extract the real and imaginary parts of a complex quantity. The Euclidean norm is denoted by $\|\cdot\|_2$, and $\tr(\cdot)$ denotes the trace. The notation $\mathcal{CN}(\mu,\sigma^2)$ denotes the circularly symmetric complex Gaussian distribution with mean $\mu$ and variance $\sigma^2$. The identity matrix of size $N$ is denoted by
$\mathbf I_N$, and the all-zero vector by $\mathbf 0_N$. Specific to this model, an underline denotes the augmented counterpart of a quantity, $\underline{\bm z}=[\bm z\transp,\bm z\herm]\transp$. 

\section{The Signal Model and Augmented Angle CRB}
\label{sec:model}

\begin{figure}[htbp]
    \centering
    \includegraphics[width=\columnwidth]{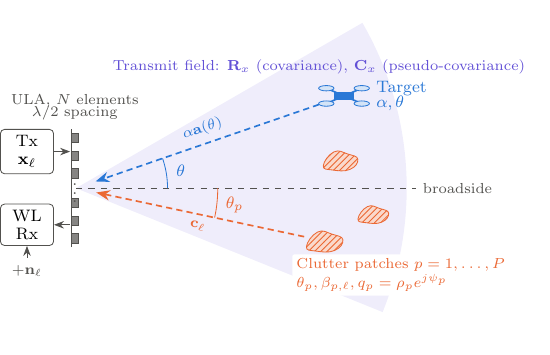}
    \vspace{-1.5em}
    \caption {System model of the considered ISAC sensing scenario, comprising an \(N\)-element uniform linear array, a point target, and \(P\) clutter patches. } 
    \label{fig:system-model}
\end{figure}

Consider that a base station, illustrated in Fig. \ref{fig:system-model}, with an $N$-element half-wavelength uniform linear array (ULA) senses a target over $L$ snapshots while transmitting $\mathbf x_\ell\in\C^N$, $\ell=1,\dots,L$, zero-mean, possibly improper and i.i.d. across snapshots, with covariance $\Rx$ and pseudo-covariance $\Cx$, defined as
\begin{equation}
\label{eq:RxCx}
\Rx=\E[\mathbf x_\ell\mathbf x_\ell\herm],\qquad
\Cx=\E[\mathbf x_\ell\mathbf x_\ell\transp].
\end{equation}
Let $\theta$ be the target angle; the effective sensing model is
\begin{equation}
\label{eq:model}
\mathbf y_\ell=\alpha\,\steer(\theta)+\mathbf d_\ell,\qquad \ell=1,\dots,L,
\end{equation}
with a steering vector $[\steer(\theta)]_n=e^{j\pi n\sin\theta}$, $n=0,\dots,N-1$, unknown complex target
amplitude $\alpha$, and a zero-mean disturbance $\mathbf d_\ell$ of covariance
$\Rd=\E\{\mathbf d\mathbf d\herm\}$ and pseudo-covariance $\Rtd=\E\{\mathbf d\mathbf d\transp\}$. For proper sensing, i.e.\ $\mathbf d_\ell\sim\mathcal{CN}(\mathbf 0,\Rd)$, $\Rtd = \mathbf 0$.

Consider a clutter region represented by $P$ discrete scattering patches, indexed by $p=1,\ldots,P$. Each patch represents a localized contribution to the received clutter, characterized by its angle of arrival $\theta_p$, steering vector $\mathbf a_p=\steer(\theta_p)$, and random reflection coefficient $\beta_{p,\ell}$. The disturbance is modeled as $\mathbf d_\ell=\mathbf c_\ell+\mathbf n_\ell$, where $\mathbf n_\ell\sim\mathcal{CN}(\mathbf 0,\sigma_n^2\mathbf I_N)$ denotes proper receiver noise, and the aggregate clutter return is
\begin{equation}
\label{eq:clutter}
\mathbf c_\ell=
\frac{1}{\sqrt N}
\sum_{p=1}^{P}
\beta_{p,\ell}\,
\mathbf a_p\mathbf a_p\transp\mathbf x_\ell.
\end{equation}

The scattering coefficients are assumed to be zero-mean and independent of the transmitted signal and receiver noise. Their second-order moments satisfy
\begin{align}
\label{eq:clutter-moments}
\E\{\beta_{p,\ell}\beta_{p',\ell}^*\}
&=\sigma_p^2\delta_{pp'},\\
\E\{\beta_{p,\ell}\beta_{p',\ell}\}
&=\sigma_p^2 q_p\delta_{pp'},
\end{align}
where $\delta_{pp'}$ is the Kronecker delta, $\sigma_p^2=\E\{|\beta_{p,\ell}|^2\}$ denotes the scattering power of patch $p$, and $q_p=\rho_p e^{j\psi_p}$, with $0\leq\rho_p\leq1$, characterizes its second-order non-circularity, and $\psi_p$ represents its pseudo-covariance phase. The illumination power $w_p$ and pseudo-variance $\widetilde w_p$ of the transmitted field incident on each patch are defined as
\begin{align}
\label{eq:illum}
w_p
&=\mathbf a_p\transp\Rx\mathbf a_p^*,\\
\widetilde w_p
&=\mathbf a_p\transp\Cx\mathbf a_p.
\end{align}

Averaging over the transmitted signal and scattering coefficients, the disturbance covariance and pseudo-covariance are respectively obtained as
\begin{equation}
\label{eq:dist-Rd}
\Rd=
\sigma_n^2\mathbf I_N+
\frac{1}{N}
\sum_{p=1}^{P}
\sigma_p^2 w_p\mathbf a_p\mathbf a_p\herm,
\end{equation}
\begin{equation}
\label{eq:dist-Rtd}
\Rtd=
\frac{1}{N}
\sum_{p=1}^{P}
\sigma_p^2 q_p\widetilde w_p
\mathbf a_p\mathbf a_p\transp.
\end{equation}

For a prescribed transmit covariance, the incident field power and, consequently, the
ordinary disturbance covariance remain fixed. In contrast, the transmit pseudo-covariance determines the incident-field pseudo-variance, thereby modifying $\Rtd$.

The target amplitude and its coherent phase reference are assumed independent of variations in $\Cx$. This formulation isolates the influence of transmit pseudo-covariance on the sensing bound from active CRB beamforming~\cite{liu2022crb}, where the transmit covariance also determines target illumination.

To account for the possible non-circularity in the sensing signal, we take advantage of widely linear processing (WLP) \cite{schreier2010complex}, a technique that jointly processes a complex signal and its complex conjugate. This allows for simultaneous exploitation of the second-order statistical information contained in the disturbance pseudocovariance, as well as its covariance. For this, we define the augmented disturbance vector as
\begin{equation}
\underline{\mathbf d}_\ell =
\begin{bmatrix}
\mathbf d_\ell\\
\mathbf d_\ell^\ast
\end{bmatrix},
\end{equation}
whose covariance is
\begin{equation}
\underline{\mathbf R}_d =
\begin{bmatrix}
\mathbf R_d & \widetilde{\mathbf R}_d\\
\widetilde{\mathbf R}_d^\ast & \mathbf R_d^\ast
\end{bmatrix}
\in\mathbb C^{2N\times 2N}.
\end{equation}

For proper disturbance, \(\widetilde{\mathbf R}_d=\mathbf 0\), and the augmented covariance becomes block diagonal. For non-circular disturbance, the off-diagonal blocks describe the additional second-order structure. Then, let $\bm\eta=[\theta,\Re\alpha,\Im\alpha]\transp$ be the real parameter vector. Writing $\mathbf a=\steer(\theta)$ and $\dot{\mathbf a}=\partial\mathbf a/\partial\theta$, the matrix of augmented mean derivatives is
\begin{equation}
\label{eq:G}
\mathbf G=
\begin{bmatrix}
\alpha\dot{\mathbf a} & \mathbf a & j\mathbf a\\
\alpha^*\dot{\mathbf a}^* & \mathbf a^* & -j\mathbf a^*
\end{bmatrix}
\in\C^{2N\times3}.
\end{equation}
For the CRB analysis, we assume independent effective complex Gaussian disturbance snapshots with the covariance and pseudo-covariance derived above. These statistics are obtained by averaging over the transmitted symbols and scattering coefficients, and are treated as known and independent of $\bm\eta$. The resulting CRB is exact for this effective Gaussian likelihood, rather than necessarily for the underlying random-product clutter distribution.

Under this assumption, the real $3\times3$ Fisher information matrix (FIM) and angle bound are
\begin{equation}
\label{eq:FIM}
\mathbf J = L\,\Re\{\mathbf G\herm\underline{\mathbf R}_d^{-1}\mathbf G\}
\end{equation}
 \begin{equation}
\label{eq:angle-crb}
\CRB(\theta) = [\mathbf J^{-1}]_{11}.
\end{equation}

We analyze specifically the first diagonal entry of the inverse FIM, corresponding to the target angle. Since the remaining parameters are the unknown real and imaginary components of $\alpha$, inverting the full FIM accounts for their uncertainty when evaluating the angle bound. By \eqref{eq:illum}, $\Cx$ determines $\widetilde w_p$, which affects $\Rtd$. Therefore, $\Cx$ affects the off-diagonal blocks of $\underline{\mathbf R}_d$. Since $\mathbf J$ depends on $\underline{\mathbf R}_d^{-1}$, as shown in \eqref{eq:FIM}, the angle CRB is, in general, a function of $\widetilde w_p$ and thus of $\Cx$.

This establishes the connection between transmit impropriety and widely-linear sensing: by modifying the pseudo-covariance of the signal-dependent clutter, the transmitter changes the second-order information available to the sensing receiver, potentially improving the angle-estimation CRB without altering the prescribed transmit covariance.

\section{Conditions for Transmit Pseudo-Covariance Sensitivity}
\label{sec:conditions}

Following the model introduced in Section~\ref{sec:model}, we now investigate the conditions under which variations in the transmit pseudo-covariance can influence the angle CRB. To obtain a tractable initial characterization for this paper, we simplify the disturbance model to a single clutter patch, $P=1$, located at angle $\theta_c$, with steering vector $\mathbf a_c=\steer(\theta_c)$, scattering power $\sigma_c^2$, and non-circularity coefficient $q_c=\rho_c e^{j\psi_c}$. The general $P$-patch analysis is left for future work.

Under this specialization, the disturbance covariance and pseudo-covariance in \eqref{eq:dist-Rd} and \eqref{eq:dist-Rtd} reduce to
\begin{equation}
\label{eq:single-Rd}
\Rd=
\sigma_n^2\mathbf I_N+
\frac{\sigma_c^2 w}{N}
\mathbf a_c\mathbf a_c\herm,
\end{equation}
\begin{equation}
\label{eq:single-Rtd}
\Rtd=
\frac{\sigma_c^2 q_c\widetilde w}{N}
\mathbf a_c\mathbf a_c\transp,
\end{equation}
where $w=w_1$ and $\widetilde w=\widetilde w_1$ denote the incident field power and pseudo-variance associated with the clutter patch, as defined in \eqref{eq:illum}.

\subsection{Projected Target--Clutter Geometry and the Projected Overlap}

The unknown complex target amplitude $\alpha$ is a nuisance parameter when estimating $\theta$, regardless of whether the FIM is ordinary or augmented. Changes that merely rescale the target response, including its overall phase, can be explained by a change in $\alpha$, so they provide no independent information about the angle. Angle information therefore comes from changes in the array response that cannot be explained by such rescaling. Such changes are captured by the orthogonal projector onto the complement of the target steering vector,
\begin{equation}
\label{eq:projector}
\mathbf\Pi
=
\mathbf I_N-
\frac{\mathbf a\mathbf a\herm}{N},
\end{equation}
where $\mathbf a=\steer(\theta)$ and $\mathbf a\herm\mathbf a=N$. The projected target steering-vector derivative and clutter steering vector are
\begin{equation}
\label{eq:projected-vectors}
\mathbf v=\mathbf\Pi\dot{\mathbf a},
\qquad
\mathbf b=\mathbf\Pi\mathbf a_c,
\end{equation}
where $\dot{\mathbf a}$ denotes the derivative of the target steering vector with respect to $\theta$. The projected target--clutter overlap is then defined as 
\begin{equation}
\label{eq:overlap}
\zeta
=
\mathbf b\herm\mathbf v
=
\mathbf a_c\herm\mathbf\Pi\dot{\mathbf a}.
\end{equation}

\begin{figure}[t]
    \centering
    \includegraphics[width=1.1\linewidth]{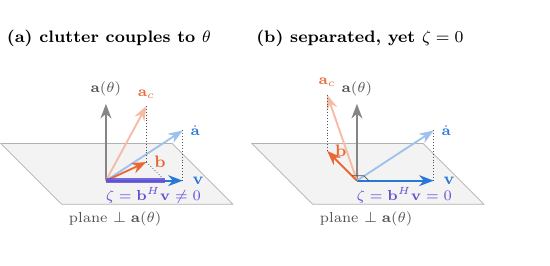}
    \vspace{-2.5em}
    \caption{Geometric interpretation of the projected target–clutter overlap. Projection onto the subspace orthogonal to the target steering vector removes the component associated with the unknown target amplitude.}
    \label{fig:target-clutter-geometry}
\end{figure}

Here, $\mathbf v$ represents the component of the target response derivative that remains informative after eliminating the unknown target amplitude, while $\mathbf b$ represents the clutter direction after the same projection. The scalar $\zeta$ therefore characterizes the geometric coupling between the clutter and the angle-informative component of the target response. When $\zeta=0$, the clutter is orthogonal to the projected angle-informative direction. Under this reduced model, the clutter does not degrade the amplitude-eliminated angle information, regardless of its pseudo-covariance. This projected overlap distinguishes angular separation from the geometric coupling relevant to angle estimation. Although distinct target and clutter directions generally produce different steering vectors, configurations with $\theta_c\neq\theta$ can still satisfy
\begin{equation}
\mathbf a_c\herm\mathbf\Pi\dot{\mathbf a}=0.
\end{equation}
In these configurations, the clutter direction is orthogonal to the amplitude-eliminated angle-sensitive component. Consequently, as shown in Fig. \ref{fig:target-clutter-geometry} angular separation alone does not establish whether transmit pseudo-covariance can influence the angle CRB.

\subsection{Conditions for CRB Sensitivity}

For a prescribed $\Rx$, the set of feasible transmit pseudo-covariances is
\begin{equation}
\label{eq:feasible-Cx}
\mathcal{C}(\Rx)=
\left\{
\Cx:\ \Cx=\Cx\transp,\
\begin{bmatrix}
\Rx & \Cx\\
\Cx^* & \Rx^*
\end{bmatrix}
\succeq\mathbf 0
\right\},
\end{equation}
which, for the single-patch illumination quantities, implies
\begin{equation}
\label{eq:illum-feasibility}
|\widetilde w|\leq w.
\end{equation}
The following theorem establishes when the angle CRB depends on the choice of $\Cx$ within this set.
\begin{theorem}
\label{thm:conditions}
Assume $\alpha\neq0$, $\sigma_n^2>0$, and $\mathbf v\neq\mathbf 0$. For a prescribed $\Rx$, the angle CRB is nonconstant over $\mathcal{C}(\Rx)$ if and only if
\begin{equation}
\label{eq:conditions-summary}
\sigma_c^2 w>0,\qquad
\rho_c>0,\qquad
\zeta\neq0.
\end{equation}
\end{theorem}

\begin{proof}[Proof of Theorem \ref{thm:conditions}]
Eliminating the unknown complex amplitude through the Schur complement of the augmented FIM in \eqref{eq:FIM} restricts the local angle information to the subspace orthogonal to $\mathbf a$, represented by $\mathbf\Pi$. Applying this projector to \eqref{eq:single-Rd} and \eqref{eq:single-Rtd} gives
\begin{equation}
\label{eq:proof-projected-moments}
\begin{aligned}
\mathbf\Pi\Rd\mathbf\Pi
&=\sigma_n^2\mathbf\Pi+
\frac{\sigma_c^2w}{N}\mathbf b\mathbf b\herm,\\
\mathbf\Pi\Rtd\mathbf\Pi^*
&=\frac{\sigma_c^2q_c\widetilde w}{N}
\mathbf b\mathbf b\transp.
\end{aligned}
\end{equation}
Thus, the projected clutter occupies only the direction $\mathbf b$. If $\mathbf b=\mathbf 0$, the clutter vanishes from this subspace and the angle CRB is independent of $\Cx$.

Otherwise, let $B=\|\mathbf b\|_2^2>0$ and define the normalized clutter direction $\mathbf e=\frac{\mathbf b}{\sqrt B}$. Projecting the angle derivative and disturbance moments onto $\mathbf e$ yields
\begin{equation}
\label{eq:proof-scalar-quantities}
\begin{aligned}
\xi
&=\alpha\mathbf e\herm\mathbf v
=\frac{\alpha\zeta}{\sqrt B},\\
r
&=\mathbf e\herm\Rd\mathbf e
=\sigma_n^2+\frac{\sigma_c^2wB}{N},\\
\widetilde r
&=\mathbf e\herm\Rtd\mathbf e^*
=\frac{\sigma_c^2q_c\widetilde wB}{N}.
\end{aligned}
\end{equation}
The resulting scalar coordinate has angle derivative $\xi$, variance $r$, and pseudo-variance $\widetilde r$. Its augmented covariance is
$\left[\begin{smallmatrix}
r & \widetilde r\\
\widetilde r^* & r
\end{smallmatrix}\right]$,
while the remaining projected coordinates contain only proper white noise. Inverting this augmented covariance and combining the information contributions gives
\begin{equation}
\label{eq:proof-angle-information}
\begin{aligned}
\CRB(\theta)^{-1}=2L\Bigg[
&\frac{|\alpha|^2\|\mathbf v\|_2^2-|\xi|^2}
{\sigma_n^2}\\
&+\frac{r|\xi|^2-\Re\{\widetilde r^*\xi^2\}}
{r^2-|\widetilde r|^2}
\Bigg].
\end{aligned}
\end{equation}

For fixed $\Rx$, if $\sigma_c^2w=0$, feasibility forces the clutter contribution to vanish. If $\rho_c=0$, then $\widetilde r=0$ for every feasible $\Cx$. Likewise, $\zeta=0$ implies $\xi=0$, eliminating the clutter-dependent information term. Hence, all three conditions in \eqref{eq:conditions-summary} are necessary.

Conversely, suppose all three conditions hold. Then $w>0$, $B>0$, and $\xi\neq0$. Feasible transmit pseudo-covariances can realize every normalized field pseudo-variance $z=\widetilde w/w$ with $|z|\leq1$; for instance, $\Cx=z\,\Rx\mathbf a_c^{*}\mathbf a_c\herm\Rx^{*}/w$ belongs to $\mathcal{C}(\Rx)$ for every $|z|\leq1$ and yields $\widetilde w=zw$. Let $m=\sigma_c^2\rho_c wB/N>0$. The feasible choices $z_\pm=\pm e^{j[2\arg(\xi)-\psi_c]}$ produce $\widetilde r_\pm=\pm m e^{j2\arg(\xi)}$. The clutter-dependent fraction in \eqref{eq:proof-angle-information} consequently takes the two values
\begin{equation}
\frac{|\xi|^2}{r+m},
\qquad
\frac{|\xi|^2}{r-m}.
\end{equation} Since $0<\rho_c\leq1$ and $\sigma_n^2>0$ ensure $r>m>0$, these values differ. Therefore, two feasible transmit pseudo-covariances yield different angle CRBs, establishing sufficiency.
\end{proof}

\subsection{The Closed-Form Optimal Transmit Pseudo-Covariance Derivation}
\label{subsec:optimal}

Building upon the aforementioned single-patch characterization, we now derive the incident-field pseudo-variance that minimizes the angle CRB at a fixed transmit covariance. Using the projected quantities from \eqref{eq:projected-vectors} and \eqref{eq:overlap}, we define
\begin{equation}
\label{eq:normalized-overlap}
\chi=\frac{|\zeta|^2}{B\|\mathbf v\|_2^2},
\qquad B=\|\mathbf b\|_2^2,
\end{equation}
and the projected clutter-to-noise ratio
\begin{equation}
\label{eq:projected-cnr}
\gamma=\frac{\sigma_c^2wB}{N\sigma_n^2}.
\end{equation}
Here, $\chi\in[0,1]$ measures the normalized projected target--clutter overlap, with the convention $\chi=0$ when $B=0$. Rewriting \eqref{eq:proof-angle-information} in terms of these quantities, the normalized angle information becomes
\begin{equation}
\label{eq:normalized-information}
\frac{f_{\mathrm{free}}}{\CRB(\theta)}
=1-\chi+\chi h,
\end{equation}
where
\begin{equation}
\label{eq:free-crb}
f_{\mathrm{free}}=
\frac{\sigma_n^2}
{2L|\alpha|^2\|\mathbf v\|_2^2}
\end{equation}
is the clutter-free bound, and
\begin{equation}
\label{eq:h-function}
h=
\frac{1+\gamma-\gamma\rho_c|z|\cos\delta}
{(1+\gamma)^2-\gamma^2\rho_c^2|z|^2},
\end{equation}
with $\delta=\arg(z)+\psi_c-2\arg(\alpha\zeta)$. For $\zeta\neq0$, minimizing the CRB amounts to maximizing $h$. Its optimal phase satisfies $\delta=\pi$, and the resulting expression $h=[1+\gamma-\gamma\rho_c|z|]^{-1}$ is maximized at $|z|=1$. Thus,
\begin{equation}
\label{eq:optimal-pseudo-illum}
\widetilde w^\star
=-w e^{j[2\arg(\alpha\zeta)-\psi_c]}.
\end{equation}
Therefore, every feasible value of $\Cx$ that satisfies this condition will yield the optimal CRB.

When $\zeta=0$, the CRB is independent of the pseudo-illumination phase. For proper signaling, $z=0$ gives $h_0=(1+\gamma)^{-1}$, whereas the optimum gives $h^\star=[1+(1-\rho_c)\gamma]^{-1}$. Substitution into \eqref{eq:normalized-information} yields the optimal relative CRB reduction
\begin{equation}
\label{eq:crb-dividend}
D_{\mathrm{CRB}}
=1-\frac{\CRB^\star}{\CRB_0}
=\frac{\rho_c\chi\gamma}
{(1+\gamma)[1+(1-\chi)(1-\rho_c)\gamma]}.
\end{equation}
Here, $\CRB_0$ and $\CRB^\star$ denote the proper and optimal improper bounds, respectively, at the same prescribed transmit covariance. 

For the edge case of rectilinear clutter, $\rho_c = 1$, $\widetilde w^\star$ completely eliminates the clutter-induced CRB penalty under the single-patch simplification. Mathematically, substituting $h^\star$ into \eqref{eq:normalized-information} yields
\begin{equation}
    \label{eq:expanded-normalized-information}
    \frac{f_{\mathrm{free}}}{\CRB(\theta)}
=1-\chi+\frac{\chi}{1+(1-\rho_c)\gamma},
\end{equation}
which, for $\rho_c = 1$, reduces to $f_{\mathrm{free}}/\CRB(\theta) = 1$, i.e., $\CRB(\theta) = f_{\mathrm{free}}$.
\section{Numerical Results}
\label{sec:validations}

In order to better comprehend the gain that can be obtained by exploiting a non-circular signal's pseudo-covariance, this section presents simulations that compare the CRB in proper and improper scenarios from two perspectives: the CRB penalty as a function of clutter--target separation, and the optimal CRB reduction as a function of clutter strength. 
For the simulations in this section, we consider a ULA of $N = 32$ elements observed over a single snapshot ($L=1$) with an isotropic transmit covariance, $\Rx = \mathbf I_N$, and a target angle of $\theta = 0^\circ$. Both comparisons also assume unit noise variance, $\sigma_n^2=1$. For the CRB penalty experiment, clutter non-circularity is fixed at $\rho_c = 0.9$, with a clutter pseudo-covariance phase of $\psi_c = 0$, target amplitude of $\alpha = 1$ and a fixed CNR of 8 dB. As the clutter separation varies, we evaluate proper CRB at $\Cx = 0$, while the optimal improper CRB is obtained through \eqref{eq:optimal-pseudo-illum}, recomputing $\widetilde{w}^\star$ for every clutter angle. A third bound is evaluated at a fixed $\Cx = 0.8\,\mathbf I_N$, and all bounds are evaluated through the augmented FIM and expressed as the clutter-induced penalty, defined as the ratio between the angle CRB and the clutter-free CRB, in dB.

\begin{figure}[t]
    \centering
    \includegraphics[width=1\linewidth]{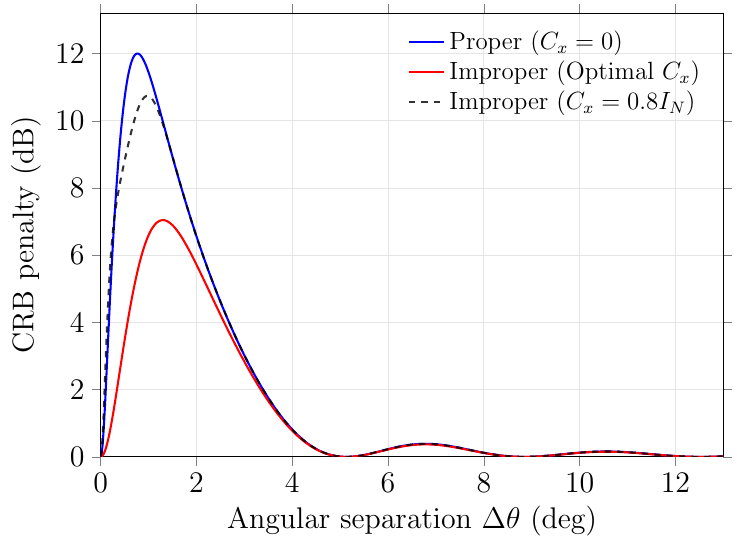}
    \caption{Angle CRB penalty versus target--clutter
    angular separation for proper signaling,
    closed-form optimal improper signaling,
    and a fixed improper design, $\Cx=0.8\,\mathbf I_N$.}
    \label{fig:crb-angle}
\end{figure}

Figure \ref{fig:crb-angle} shows the CRB penalty as a function of clutter--target separation. Results show that, in all evaluated scenarios, the CRB penalty is largest when $0^\circ< \Delta\theta < 2^\circ$, i.e., when clutter and target are close together but not coincident. The penalty reduction, however, is substantial for the optimized improper design, which achieves up to approximately 7~dB lower penalty than the proper case.

The second experiment quantifies the maximum improvement of optimal improper sensing relative to proper sensing, under the same conditions, using \eqref{eq:crb-dividend}. The projected clutter-to-noise ratio $\gamma_{\mathrm{dB}}=10\log_{10}\gamma$ is varied from $-10$ to $40$ dB and curves are plotted to represent this dividend for given values of overlap and circularity. 

\begin{figure}[htb]
    \centering
    \includegraphics[width=1\linewidth]{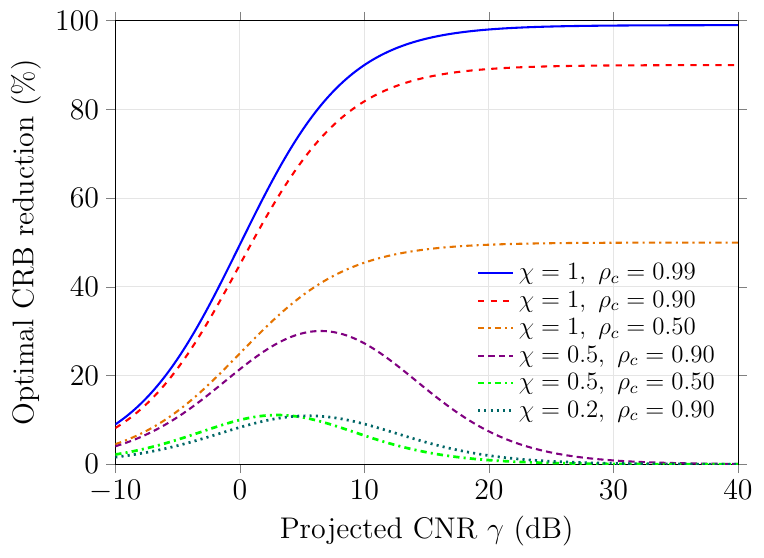}
    \vspace{-1.5em}
    \caption{Optimal relative angle CRB reduction versus projected clutter-to-noise ratio for different normalized overlap factors $\chi$ and clutter non-circularity coefficients $\rho_c$.}
    \label{fig:crb-dividend}
\end{figure}

Figure~\ref{fig:crb-dividend} presents the optimal CRB reduction, $100D_{\mathrm{CRB}}$, relative to proper signaling. For complete projected overlap ($\chi=1$), the reduction increases with clutter strength, approaching $99\%$, $90\%$, and $50\%$ for $\rho_c=0.99$, $0.90$, and $0.50$, respectively. At $\rho_c=0.90$, reducing the overlap to $\chi=0.5$ limits the maximum CRB reduction to approximately $30\%$, attained near $6.5$ dB, while $\chi=0.2$ yields a maximum of approximately $11\%$ near $5$ dB. Unlike the complete-overlap cases, these partial-overlap curves decrease toward zero at high clutter-to-noise ratios. These results demonstrate that strong clutter non-circularity alone does not guarantee substantial sensing gains, as the achievable improvement also depends on the projected target--clutter overlap.

These results illustrate how the three conditions established in Section~\ref{sec:conditions} determine whether impropriety can affect the angle CRB, while the projected overlap, clutter strength, and non-circularity determine the magnitude of the achievable reduction.
\section{Conclusion}
\label{sec:conc}

In this work, we have derived the necessary and sufficient conditions under which the angle CRB of an ISAC system can be affected by the transmit pseudo-covariance. Using widely linear processing, we defined an augmented multi-patch clutter sensing signal model and analytically described how the angle CRB is affected by the signal's pseudo-covariance through the pseudo-variance of the incident field. Then, using a single clutter patch model, we established the necessary and sufficient conditions required for the pseudo-covariance to change the CRB, and derived a closed-form optimal transmit pseudo-covariance that minimizes the CRB under these conditions. We then utilized the defined equations to verify the paper's findings in simulations comparing CRB penalty with regards to target-clutter separation in proper and improper scenarios. In this context, all scenarios demonstrate a high CRB penalty at small angular separations, while the optimally improper design achieved up to approximately 7~dB lower penalty in the same separation window. Using the same optimal derivation, we compared the improper CRB sensing gain relative to proper sensing in different degrees of projected overlap and non-circularity. For complete overlap, the optimal transmit pseudo-covariance achieves up to 99\% relative reduction, and eliminating the clutter-induced CRB penalty in the case of rectilinear clutter. Reducing the projected overlap limits the relative gain to at most 30\%. Future work includes further development of adequate improper conditions in a multi-clutter patch scenario, as well as the expansion of this concept to the CRB-rate tradeoff issue.

\balance

\bibliographystyle{IEEEtran}
\bibliography{refs}

\end{document}